# Solution-dependent electrostatic spray deposition (ESD) ZnO thin film growth processes

Fysol Ibna Abbas[1,2,3]* and Mutsumi Sugiyama[1,4]

[1] Department of Electrical Engineering, Tokyo University of Science, 2641 Yamazaki, Noda, Chiba 278-8510, Japan
[2] Department of Theoretical Physics, University of Dhaka, Dhaka-1000, Bangladesh
[3] Department of Electrical and Electronic Engineering, City University, Dhaka 1216, Bangladesh
[4] Research Institute, RIST, Tokyo Univ. of Science, E-mail: fysolibnaabbas1988@gmail.com

**Abstract**

The present study describes a facile route of zinc oxide (ZnO) grows using the solution-dependent electrostatic spray deposition (ESD) method at temperatures ranging from 300 °C - 500 °C. In this work, zinc chloride ($ZnCl_2$) was dissolved in ethanol ($CH_3CH_2OH$) to prepare the 0.1 M concentration of 20 ml for spray solution by ESD. Adding different deionized water ($H_2O$) ratio, three different solutions were prepared. The results reveal that adding $H_2O$ ration, suppressing the c-axis crystal growth of ZnO thin films. The adhesion of anions was believed to be responsible for this suppression. XRD texture analysis examined the preferred orientations of the (100) and (002) planes of the ZnO thin films. Microstructural parameters namely, lattice parameters, bond length, positional parameters, full width at half maximum, crystallite sizes, lattice strain, and lattice dislocation density, are investigated. This research marks a turning point in cost-effective industrial and commercial applications for ESD-deposited electronics.

* E-mail: fysolibnaabbas1988@gmail.com

# 1. Introduction

Zinc oxide (ZnO) nanostructures have been the focus of significant research because their potential for use as a material in various fields, including solar cells [1], light-emitting diodes [2,3], photocatalysis [4], gas sensors [5], catalysis [6,7], laser diodes [8], varistors [6,9] and sensors [10], is promising. Zinc oxide (ZnO) is a widely recognized n-type semiconductor with an enormous band gap of approximately 3.37 (eV) and a substantial exciton binding energy of around 60 (meV) at ambient temperature [11]. The influence of structural morphology, crystal size, and crystalline density on the characteristics of nanostructured ZnO has been firmly established [12]. So far, numerous types of ZnO nanostructures, including nanowires [13], nanotubes [14], nanorods [15], and nanodisks [16], have been developed using different techniques such as vapor phase deposition [17], thermal oxidation [18], electrochemical methods [19], and hydrothermal procedures [20-25].

However, the use of the methodologies for ZnO nanostructures is restricted due to factors such as elevated temperature, precise gas concentration, hazardous chemical reagents, and expensive equipment. The current goal is to create a simple and cost-efficient method that can be easily expanded for commercial use. Well-established methods for depositing ZnO thin films include dry processes like sputtering [26-34] and pulsed laser deposition [35,36], as well as wet processes like chemical bath deposition [37,38], mist chemical vapor deposition [39], and spray pyrolysis [40-42]. Additionally, the electrostatic spray deposition (ESD) method is not only easy to use, but it also has important industrial benefits, such as the ability to cover large areas because it is simple and easy to understand. This technique allows for non-vacuum deposition, has a low cost, and provides convenient control over the composition ratio and doping [43-50].

The ESD approach is anticipated to be employed to produce cost-effective and uncomplicated electronics that are transparent to visible light, owing to these benefits. Currently, there are a limited number of research works and their corresponding papers that specifically address the growth mechanism of oxide-based semiconductors utilizing the ESD approach [51] and focus on the particular solution utilized. Characterization using the ESD approach is essential for comparing the results with other techniques. Furthermore, the origin of the oxygen atoms in the growth process of ZnO thin films is also investigated in the presence of $H_2O$ solutions.

This study aimed to investigate the ZnO growth mechanism by ESD, varying the $H_2O$ ratio in the precursor solutions in different temperature ranges, while considering the issue of cost-effectiveness. In this case, 20 ml of three different solutions were prepared by changing the $H_2O$ ration for the ESD technique. Zinc chloride ($ZnCl_2$) is used as a starting material and a source for Zinc, which is dissolved in the prepared 0.1 M ethanol ($CH_3CH_2OH$). The ESD method is applied at three different temperatures, 300 °C, 400 °C, and 500 °C, on an ITiO glass substrate to investigate how ZnO grows using the ESD method, respectively. Furthermore, nanoscale-level studies mainly lattice parameters, a (Å), and c (Å) [52,53], bond length, (Zn-O bond (Å)) [54,55], and, positional parameter, µ, [54] were investigate to understand the preferential growth of ZnO thin films and how behavior is changing and affected by ESD with changing the $H_2O$ ration. The average crystallite size, D (Å) [56,57] was calculated from the XRD peak width of (002) based on the Debye–Scherrer equation (8) [58]. Peak width analysis yields two primary properties, namely crystallite size, D (Å), and lattice strain, ε (%) [60]. So far, these microstructural parameters were also investigated for the ZnO thin film growth mechanism by a novel ESD technique. The equations are involved to calculate these parameters are described in the results and discussion section. These works represent the initial phase in the development of affordable and uncomplicated devices for the oxide-based semiconductor that are transparent to visible light. It also aims to investigate ZnO growth processes using the ESD approach.

## 2. Experimental methods

The zinc sources used in this study was zinc chloride ($ZnCl_2$ (98% Assay, $ZnCl_2$=136.32), Lot. LEH7398, Mfg. Date. 2021.09, FUJIFILM Wako Pure Chemical Corporation). The solutes were ethanol ($CH_3CH_2OH$ (99.5), Cat. No. 14033-70, FW: 46.07, KANTO CHEMICAL CO, INC), and deionized water ($H_2O$). ZnO films were deposited using the electric field applied spray pyrolysis named electrostatic spray deposition (ESD) on a conductive $In_2O_3$:Sn (ITiO) coated alkali-free glass substrate [51]. The experimental diagram of ESD is shown in Fig. 1 (a-d). 20 ml of three different spray solutions were prepared with changing the $H_2O$ ration for the ESD spray technique. For preparing the first ESD solution with 0.1 M concentration, 20 ml of $CH_3CH_2OH$ was taken in a glass beaker. The density of $CH_3CH_2OH$ was 0.78945 g/cm$^3$. Then 0.2726 g $ZnCl_2$ was added with the $CH_3CH_2OH$ in the glass beaker. For making the solution homogeneous, it was heated up to 20 minutes in the magnetic stirrer with hot plate (ADVANTEC SR/350) (Fig-1 (c)). The temperature was considered 150 ºC for the magnetic stirrer with hot plate. Two separate glass beakers were used to prepare the second ESD spray solution, which consisted of 80% $CH_3CH_2OH$ (16 ml) and 20% $H_2O$ (4 ml), respectively. Next, the same amount of $ZnCl_2$ were added to the glass beaker along with 16 ml of $CH_3CH_2OH$. The same procedure was applied to prepare the homogeneous spray solution. Then 4 ml of $H_2O$ was poured into the solution containing $CH_3CH_2OH$ and $ZnCl_2$. The mixture of the solution was heated to an additional heating period of 10 minutes using a magnetic stirrer with a hot plate. For the third ESD spray solution, 50% $CH_3CH_2OH$ (10 ml of $CH_3CH_2OH$) and, 50% $H_2O$ (10 ml of $H_2O$) were taken in two different glass beakers. The same amount of $ZnCl_2$ was added with the $CH_3CH_2OH$ in the glass beaker. For making the homogeneous solution, a similar technique was applied. After 20 minutes, 10 ml of $H_2O$ was added to the $CH_3CH_2OH$ and $ZnCl_2$ solutions and heated again for 15 minutes in the magnetic stirrer with a hot plate.

ESD spray was performed for 5 minutes at a solution flow rate of 2.0 ml/h on the substrate surface, and an applied voltage between nozzle and substrate of 8 kV during spraying. Followed by ESD temperature range from 300 ºC, 400 ºC, and 500 ºC in air for 5 minutes. For the simplification and understanding the ESD technique, the parameters are listed in Table.1. of this investigation. The crystallographic orientations of the ZnO films were evaluated using X-ray diffraction (XRD; Rigaku Ultima IV, Rigaku SmartLab). Several microstructural parameters, lattice parameters, $a$ (Å), and $c$ (Å) [52,53], positional parameter, $\mu$ [54], and bond length, (Zn-O bond (Å)) [54,55], crystallite size, D (Å) [56,57], lattice strain, $\varepsilon$ (%) [60], dislocation distribution, $\delta$ (*$10^{18}$ m$^2$) [60] have been studied detailed which is correlated to the basic crystal growth [61] mechanism by the novel ESD method. The equations are involved to calculate these parameters are described in the results and discussion section.

# 3. Results and discussion

## 3.1. The XRD analysis

XRD XRD experiment has been performed to characterize the structural properties of the ZnO thin film growth mechanism using the ESD technique for 300 °C, 400 °C, and 500 °C temperatures, respectively. Figure 2 (a-b) represents the XRD pattern of the ZnO thin films on conductive $In_2O_3$:Sn (ITiO) coated alkali-free glass substrate. In the precursor solution, the $H_2O$ ratio is changed to see the effect of the ZnO growth mechanism by ESD. In Figure 2 (a), it shows a prominent, strong peak at the diffraction angle around 34.50° for first ESD spray solution sample which possesses $ZnCl_2$ and $CH_3CH_2OH$ of ZnO thin film corresponding to the (002) plane at 400 °C and 500 °C temperature, respectively. Other peaks are also appeared for (100), (101), (102) plane. The peak for 300 °C temperature at the (002) plane is found very low in Fig 2(a). However, the peak for the temperature 300 °C at the (100) plane is found little stronger compare to 400 °C. In addition, the strong peak for the (101) plane is only appeared for the 500 °C temperature of the 0% $H_2O$ by the ESD technique. A low peak strength for the (102) plane was also observed, which shifted to the left with increasing temperatures. Moreover, the consecutive peak appearance of the (002) plane in the observation, corroborated the fabrication of ZnO thin films with a hexagonal wurtzite structure [62]. In Figure 2 (b), it shows the strong peak for (002) plane at 400 °C and 500 °C temperature, respectively. But very weak peak was observed for 300 °C temperature on this plane. It was observed that the 400 °C temperature's peak strength was weaker than the 500 °C temperature; in addition, the 500 °C temperature's peak position shifted slightly to the right, suggesting a decrease in the lattice constant along the c-axis at high temperatures. The peak strength for the (100) plane was found to be weak at temperatures ranging from 300 °C to 500 °C. The other peak positions corresponding to ZnO thin film growth on the ITIO glass substrate were also observed at the (101) and (102) planes for the 20% of $H_2O$ mixed ESD spray solution. Figure 2 (c), illustrated the xrd results for 50% of $H_2O$ mixed ESD spray solution. Similar, preferred orientation of ZnO thin films was also observed for (100), (002), and (102) plane. The peak strength for (102) plane was found comparatively strong with other planes with increasing the temperatures. The peak strength on (002) plane was found gradually decreasing with increasing the $H_2O$ ration. The growth of ZnO thin films in the c-axis orientation is determined by the fundamental theory of crystal growth [63]. This theory asserts that the (002) plane of ZnO possesses the lowest surface energy, which plays a crucial role in determining the crystal structure forms [57,63]. All samples have lines corresponding to the (1 0 0), (0 0 2), (1 0 1), and (1 0 2) planes. As ESD temperatures rose, diffraction peaks narrowed and intensified. This suggests the creation of a crystalline structure. The XRD measurements confirm that all the deposited thin films exhibit a significant orientation along the c-axis (002), with variations in peak strength.

## 3.2. ESD temperature effect on lattice parameters

ZnO crystallizes in the wurtzite structure, where oxygen atoms are distributed in a, hexagonal close-packed pattern and zinc atoms occupy half of the tetrahedral positions.

The Zn and O atoms exhibit tetrahedral coordination with each other, resulting in an identical position. The Zn structure is characterized by an open configuration, where all the octahedral sites and half of the tetrahedral sites are unoccupied. According to Bragg's law [64], where $n$ is the order of diffraction (usually n=1), $\lambda$ is the X-ray wavelength and $d$ is the spacing between planes of given Miller indices h, k and l. In the ZnO hexagonal structure, the plane spacing

$$n\lambda = 2d\cos\theta \tag{1}$$

$d$ is related to the lattice constants a, c and the Miller indices by the following relation [64],

$$\frac{1}{d_{hkl}^2} = \frac{4}{3}\left(\frac{h^2+k^2+hk}{a^2}\right) + \frac{l^2}{c^2} \tag{2}$$

Considering the first-order approximation, $n$ =1, equation (2) can be written:

$$\sin^2\theta = \frac{\lambda^2}{4a^2}\left[\frac{4}{3}\left(\frac{h^2+k^2+hk}{a^2}\right) + \frac{a^2}{c^2}l^2\right] \tag{3}$$

Following the equation (3), the lattice constant $a$ for the (100) plane is calculated by [52,53]

$$a = \frac{\lambda}{\sqrt{3}\sin\theta_{100}} \tag{4}$$

Similarly, for the (002) plane, the lattice constant $c$ is calculated by the equation (3), [52,53],

$$c = \frac{\lambda}{\sin\theta_{002}} \tag{5}$$

The calculated lattice parameters $a$, and $c$ using equation (4) and (5) for this study obtained from ESD is plotted in Fig. 3(a-i). It is well known that the lattice parameters are temperature dependent, i.e. an increase in temperature leads to expansion of the lattice [65]. In Figure 3. (a-c), showed the changing effect of lattice parameter, $a$ for the (100) plane, Fig. 3. (d-f) lattice parameter, $c$ for the (002) plane; and Fig. 3 (g-i) ratio of lattice parameter, c/a by the ESD samples with changing the water ration and the temperature range from 300°C to 500°C. The lattice parameter analysis results showed a very consistent change in behavior.

In Fig. 5(a), the lattice parameter, $a$, is slightly decreasing from 300 °C to 400 °C temperature and then it starts to increasing from 400 °C to 500 °C temperature. But, the nature of lattice parameter, c is increasing from 300 °C to 400 °C, and then very constant increasing behavior is observed from 400 °C to 500 °C in Fig. 5(d). This result represents for the ESD sample without the $H_2O$ mixing in the spray solution. In Fig. 5(b), the second ESD spray solution which contains the 20% $H_2O$ mixing, the lattice parameter, $a$, is increasing from 300 °C to 500 °C. But, the nature of lattice parameter, c is increasing from 300 °C to 400 °C, and then decreasing behavior is observed from 400 °C to 500 °C in Fig. 5(e). In the third ESD spray solution which contains the 50% $H_2O$ mixing ratio, the lattice parameter, $a$, is increasing from 300 °C to 400 °C temperature and then it starts to decreasing from 400 °C to 500 °C temperature in Fig. 5(c). On the other hand, the lattice parameter, c, is slightly decreasing from 300 °C to 400 °C temperature and then it's very sharp decreasing behavior is found from 400 °C to 500 °C temperature in Fig. 5. (f). The lattice parameter ratio, c/a is plotted in Fig. 5. (g-i). The similar characteristic behavior is observed for the lattice parameter, a, and c.

### 3.3. ESD temperature effect on lattice bond length (Zn-O)

The other microstructure parameters namely positional parameter (μ), and (Zn-O) (Å) bond length correlated with the lattice parameter, a, and, c is also studied; and they are plotted in Fig. 4 (a-c) and Fig. (d-f), respectively. The equation has used to estimate the bond length (Zn-O) for the ZnO thin films is,

$$L = \left( \frac{a^2}{3c^2} + (0.5-\mu)^2 * c^2 \right)^{1/2} \tag{6}$$

where (μ) is the positional parameter of the wurtzite structure that indicates the extent of atoms displacement relative to the following plane in the c axis, as expressed with equation (7) [28,29],

$$\mu = \frac{a^2}{3c^2} + 0.25 \tag{7}$$

The detailed feature of the positional parameter, μ is illustrated in Fig. 4(a-c). For 0% and, 20 % $H_2O$ ration ESD sample showed decreasing nature from the temperature range 300 °C to 400 °C. From 400 °C to 500 °C temperature it showed the increasing nature. The 50% $H_2O$ ration ESD sample showed the increasing nature from 300 °C to 500 °C. In Fig. 4(d), the 0% $H_2O$ ration ESD sample, the nature of bond length, (Zn-O) (Å), has found the very small decreasing from 300 °C to 400 °C temperature and then it starts to increasing from 400 °C to 500 °C temperature. This very similar to the lattice constant, a in the (100) plane. In Fig. 4(e), for 20% $H_2O$ ration ESD sample, the bond length, (Zn-O) (Å) increasing from 300 °C to 400 °C temperature and then very constant nature found from 400 °C to 500 °C temperature. This is due to the decreasing the lattice constant parameter, c in this temperature range. The results for the 50% $H_2O$ ration ESD sample showed the decreasing nature from 300 °C to 500 °C temperature. This is due to the both the lattice constant parameter, a, and c are found decreasing in nature except for the 300 °C to 400 °C temperature of lattice constant parameter, a. The calculated ZnO bond length (Zn-O) (Å) values (Fig. 4. (d-f)) are in excellent agreement with the 1.98 A° reported in literature [28,29].

### 3.4. ESD temperature effect on full width at half maximum (FWHM)

XRD peak profile analysis is a simple and powerful method to evaluate the peak broadening with crystallite size and lattice strain due to dislocation. The breadth of the Bragg peak is a combination of both instrument and sample dependent effects. For an accurate analysis for size and strain effects, the instrumental broadening must be accounted. Figures 5 (a-c), and 5 (d-f) illustrate the impact of ESD technique for ZnO thin films growth mechanism on the full width at half maximum (FWHM) in the precursor spray solution for the (100) and (002) planes, respectively, at temperatures of 300°C, 400°C, and 500°C, while varying the $H_2O$ ratio. Both plane analyses for the 0% of the H₂O mixed solution show (Fig. 5. (a,d)) a low FWHM value at 300°C. With increasing temperature, the FWHM value also increases up to 500°C. However, the result indicates that a ZnO thin film is being formed on the ITiO substrate Fig. 2. (c) . From the XRD data analysis, it is observed that the corresponding peaks for the ZnO thin film are well in agreement. Despite the weak peak strengths for 300°C, the study under investigation found the patterns that are similar to those from the previous ESD study related to growth mechanisms for semiconductor oxide material [51]. Fig. 5. (b), and Fig. (e) represent the results for FWHM of (100) plane and (002) plane respectively. In Fig. 5. (b), the FWHM value is increasing from 300°C to 400°C and then decreasing from 400°C to 500°C temperature. This is correlated with the increasing the lattice constant parameter a (Å) (Fig. 3. (b)). In addition, the FWHM value is increasing from 300°C to 500°C temperature for the (002) plane. Due to this reason, the lattice constant parameter c (Å) (Fig. 3. (e)) decreasing after 400 °C temperature. The FWHM is observed in increasing in nature for (100) plane and (002) plane in Fig. 5. (c) and, Fig. 5 (f), respectively. Due to this reason, the decreasing nature of lattice constants parameter a (Å) and c (Å) decreased in the for the 50 % H₂O ration ESD spray solution.

## 3.5. ESD temperature effect on crystallite

It is known that a perfect crystal would ideally continue infinitely in all directions, but in reality, all crystals are imperfect since they have a limited size [59]. The broadening of diffraction peaks in materials occurs due to the deviation from perfect crystallinity. Crystallite size, D (Å) refers to the dimensions of a domain that exhibits coherent diffraction. It is important to note that the crystallite size of particles is typically different from their particle size because of the existence of polycrystalline aggregates. Lattice strain, ε (%) is a measure of the distribution of lattice constants ($a$ (Å), and $c$ (Å)) arising from crystal imperfections, such as lattice dislocation, δ (*$10^{18}$ m²). The X-ray line broadening is used for the investigation of dislocation distribution, δ (*$10^{18}$ m²) [60]. In the present study it was found that  the ZnO (002) plane diffraction peak is much stronger than the ZnO (101) peak (Fig. 2. (a-c)). This indicates that the formation of ZnO nanocrystals have a preferential crystallographic (002) orientation. The average crystallite size was calculated from XRD peak width of (002) based on the Debye–Scherrer equation [58],

$$D = \frac{K \lambda}{\beta \cos\theta} \qquad (8)$$

where $\beta$ is the integral half width, K is a constant equal to 0.90, $\lambda$ is the wave length of the incident X-ray ($\lambda$ = 0.1540 nm), $D$ is the crystallite size, and θ is the Bragg angle. The average crystallite size calculated for synthesized ZnO nanoparticles was 48.0~49.0 nm by ESD study. The crystallite size is assumed to be the size of a coherently diffracting domain and it is not necessarily the same as particle size. The measured crystallite size, D (Å) (Fig. 6. (a-c)) of the films is observed is similar behavior compare to lattice constant parameter c (Å) (Fig. 3. (a-c)) from temperature 300°C to 500°C. This behavior is consistent and support the growth mechanism of ZnO thin film growth mechanism [66] by the ESD technique.  The dislocation density (δ) is calculated using equation (9) [60]

$$\delta = \frac{1}{D^2} \tag{9}$$

which is described as the length of dislocation in the material per unit volume; however, the dislocation is a line defect. As the temperature increases, the dislocation density (δ) tends to decrease for 0% H₂O mixture solutions in Fig. 6. (d) from the temperature from 300°C to 500°C. The decreasing is found very small from 400°C to 500°C. But, in Fig. 6. (e), the 20% H₂O mixture ESD spray solution shows the decreasing behavior from the temperature 300°C to 400°C, and it starts to increase from 400°C to 500°C. The 50% H₂O mixture solution showed a small increase from the temperature 300°C to 400°C, while from 400°C to 500°C it increases very rapidly. Although the values of dislocation density (δ) are very small in change, but the lower values of dislocation density are attributed due to the lower level of defect and better crystalline quality of the deposited films by the ESD.

The lattice strain (ε) of the deposited ZnO thin films of different thicknesses has been calculated

$$\varepsilon = \frac{\beta \cos\theta}{4 \sin\theta} \tag{10}$$

using the equation (10) [60], where $\beta$ is FWHM (in radian) and $\theta$ indicates the angle of diffraction. The obtained values of lattice strain, ε are plotted in Fig. 6. (g-i). In Fig. 6. (g), the 0% H₂O mixed solutions showed the increasing nature from the temperature range 300°C to 500°C where, the increasing rate is found very slow from 400°C to 500°C. On the other hand, the 20% and, 50% H₂O mixed solutions showed the decreasing in nature from the temperature 400°C to 500°C in Fig. 6. (h) and, Fig. 6. (i), respectively. But, slightly, increasing nature is observed from temperature range 300°C to 400°C for 20% H₂O mixed solutions. This changing behavior of lattice strain, ε for the ZnO thin film growth mechanism indicate that increasing the H₂O ration in the solution Decreasing the lattice strain, ε by ESD technique. This could be considered an interesting feature for the ESD method regarding the oxide-based semiconductor growth technique.

### 3.6 Raman spectroscopy

Raman spectroscopy can provide insights into several molecular phenomena in a substance, including the identification of multiple phases within the same material. Moreover, it facilitated the examination of transitions from amorphous to crystalline phases, diverse flaws, and stress conditions. The current experiment demonstrated the development of thin ZnO films with a hexagonal wurtzite structure. The literature [67] states that ZnO is a semiconductor material with has $c_{6v}^4$ spatial symmetry. Group theory states that ZnO has a total of four atoms in each unit cell for its hexagonal wurtzite structure. This arrangement results in 12 phonon branches, with nine being optical modes and three being acoustic modes. Fig. 7 shows the Raman spectrum of the ZnO thin films. The Raman peaks around 475 cm⁻¹ and 780 cm⁻¹ were assigned to ZnO E2 (high) and L1 longitudinal optical (LO) modes, respectively. Raman spectroscopy of the E2 phonons plays a significant role in the study of residual stress within ZnO crystals because the stress induced in wurtzite structure crystals affects the E2 phonon frequency, hence, allowing the extraction of information on stress from the E2 mode [68,69]. A decrease in the E2 phonon frequency is attributed to tensile stress while its increase is attributed to compressive stress [68,69]. The study under this investigation, the E2 vibration mode at 475 cm$^{-1}$ is characteristic of the wurtzite phase and its value is higher than 439 cm$^{-1}$ for the stress-free bulk ZnO [70,71], suggesting that our ZnO thin films were under high tensile stress. The stress probably originated from a mismatch in the thermal expansion coefficient of the ZnO thin film (4.75 × 10$^{-6}$ K$^{-1}$) and the glass substrate (2.60 × 10$^{-6}$ K$^{-1}$) [70]. The relatively higher intensity and sharp peak of the E2 (high) mode as compared to the other peaks also indicated that the ZnO thin films had a hexagonal wurtzite phase with polycrystallinity. This was consistent with XRD analysis. The A1 (LO) mode at 780 cm$^{-1}$ originates from defects such as oxygen vacancies and Zn interstitials [71] and its relatively low intensity peak indicates a relatively low density of defects in the ZnO thin films. This corroborates the experimental XRD analysis by the ESD technique. In addition, the position of the A1 (LO) mode obtained in this study was comparatively high with that of bulk ZnO (574 cm$^{-1}$) [70,71]. The weak peak at 260 cm$^{-1}$ was attributed to second order Raman processes [70,71].

## 4. Conclusions

In conclusion, the ESD method on a conductive In$_2$O$_3$:Sn (ITiO)-coated alkali-free glass substrate has investigated the solution-dependent growth mechanism of ZnO semiconductor thin film. According to XRD, the crystallinity of the films increases by an order of 10$^{-2}$ as the temperature rises. Additionally, the XRD results demonstrated the growth of ZnO thin films with a hexagonal wurtzite structure. The lattice parameter, *a,* and *c,* analysis confirm the growth of ZnO thin film by the novel ESD technique. Several microstructure parameters were also studied to reveal the ZnO growth mechanism from the microstructure point of view point by ESD. Fair correlations were observed among the bond length (Zn-O bond), positional parameter (μ), lattice strain (ε), and lattice dislocation density (δ). The ZnO E2 (high) and A1 (LO) Raman modes were observed at 475 cm$^{-1}$ and 780 cm$^{-1}$, respectively. This study found that ESD can develop the oxide-based semiconductor thin film technique cost-effectively after analyzing all these parameters related to solution-dependent crystal growth mechanisms. Importantly, future studies on thin-film semiconductor growth mechanisms may apply using the ESD method. Altogether, these findings indicate that employing ESD for depositing the ZnO film is advantageous for creating stacked thin film devices. This work is important as it is the primary stage in developing cost-effective and uncomplicated devices that are transparent to visible light using the ESD method for semiconductor oxide-based materials. It also aims to investigate the ZnO growth process using ESD.

**Acknowledgements**

The authors express their sincere appreciation to Mr. Keisuke Tomono, Mr. Keito Okubo, and Mr. Kohta Hori for their support in completing the experiments and for the constructive discussions. We also gratefully thank MEXT for the scholarship of supporting international student (Fysol Ibna Abbas), JSPS KAKENHI (Grant number: 21K04696), and the Renewable Energy Science and Technology Research Division of the Tokyo University of Science.

## Figure captions

Fig. 1. Schematic diagram of ESD.

Fig. 2. Representative XRD patterns of ZnO film according to weight ratio of H$_2$O/solvent in the solution deposited by ESD.

Fig. 3. The trend of (a-c) lattice parameter, a; (d-f) lattice parameter, c; (c) ratio of lattice parameter, c/a, of ZnO film according to weight ratio of H$_2$O/solvent in the solution deposited by ESD at 300 °C, 400 °C and 500 °C respectively.

Fig. 4. The trend of (a-c) positional parameter, μ, and (d-f) Zn-O bond length of ZnO film according to weight ratio of H$_2$O/solvent in the solution deposited by ESD at 300 °C, 400 °C and 500 °C respectively.

Fig. 5. FWHM for the (a-c) (100) plane, and (d-f) (002) plane of the XRD peak according to weight ratio of H$_2$O/solvent in the solution deposited by ESD.

Fig. 6. The trend of (a-c) lattice crystallite, D; (d-f) Lattice dislocation density δ, (g-i) Lattice strain, ε of the XRD peak of ZnO film according to weight ratio of H$_2$O/solvent in the solution deposited by ESD at 300 °C, 400 °C and 500 °C respectively.

Fig. 7. Raman spectra of the ZnO thin films of ZnCl$_2$ and CH$_3$CH$_2$OH precursor solution by ESD at 500°C temperature ESD sample with micro-ring structures.

**Table 1. Typical deposition parameters and conditions for ESD**

| S. no. | Parameter | values |
|---|---|---|
| 1. | Distance between the nozzle aperture and the hot plate | 3.5 cm |
| 2. | Flow rate | 2.0 ml/h |
| 3. | Deposition time | 5 min |
| 4. | Applied voltage | 8 kV |

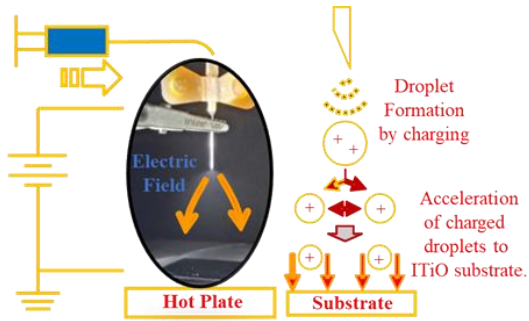
(a). Schematics diagram of ESD

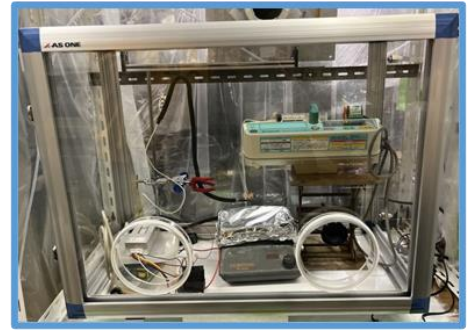
(b). Experimental configuration of of ESD

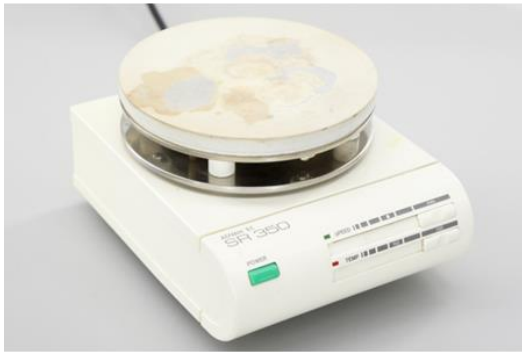
(c). Magnetic stirrer with hot plate
(ADVANTEC SR/350)

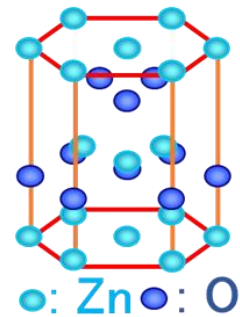
(d). Hexagonal wurtzite structure of ZnO

Fig. 1. Schematic diagram of electrostatic spray deposition (ESD).

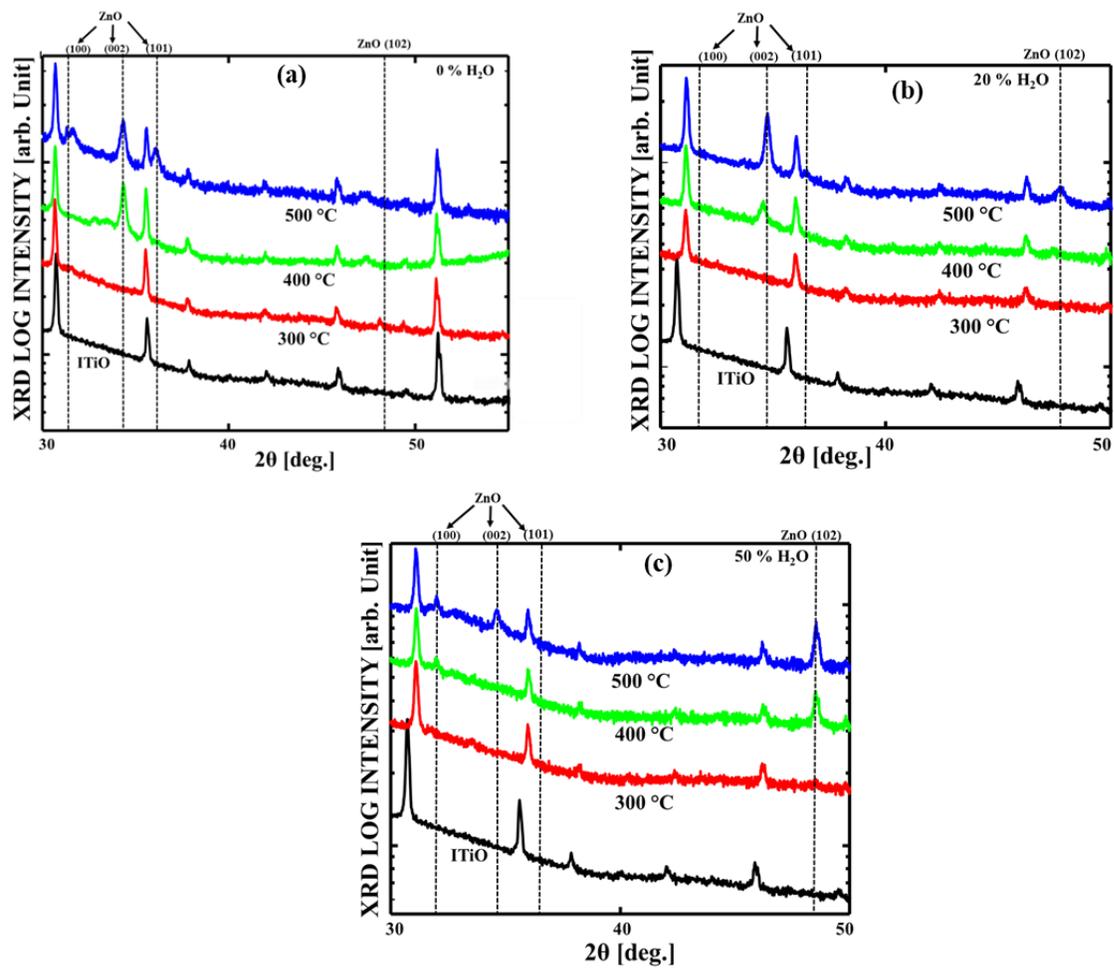

Fig. 2. XRD patterns of ZnO film according to weight ratio of $H_2O$/solvent in the solution deposited by ESD.

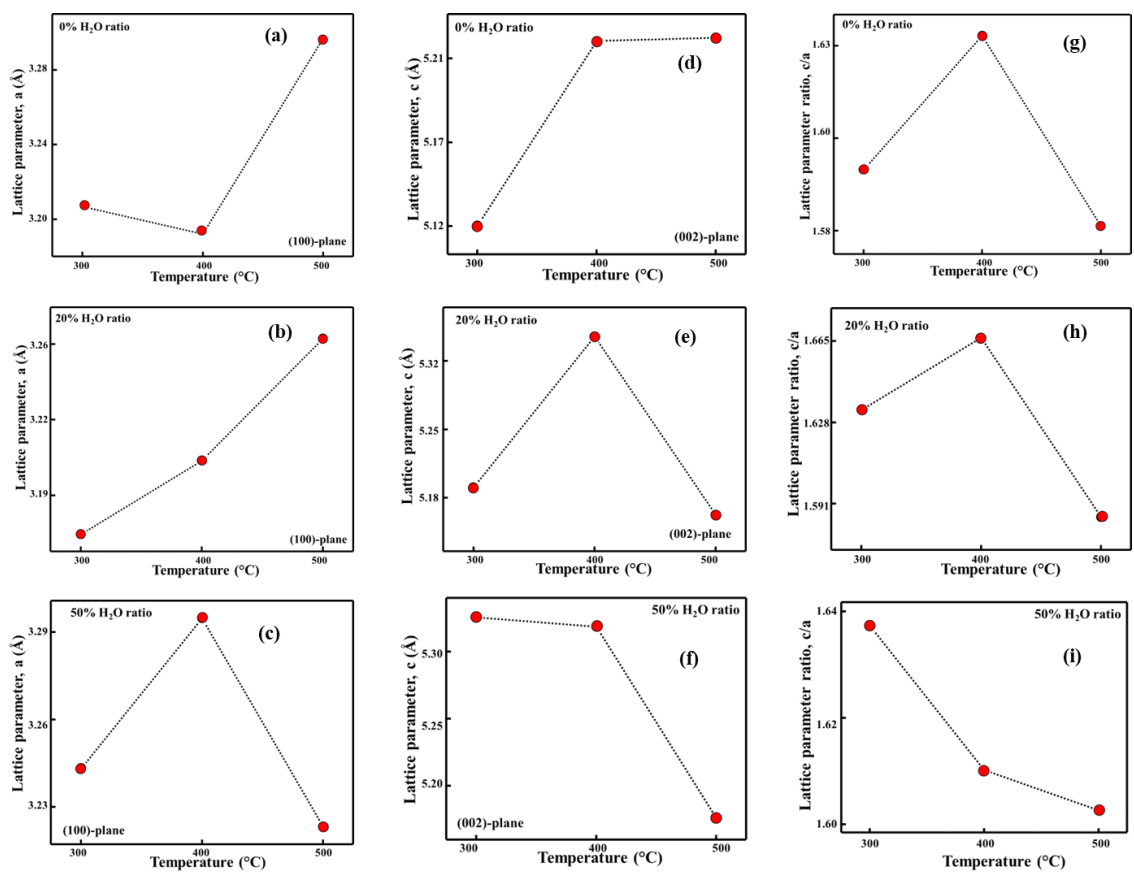

Fig. 3. The trend of (a-c) lattice parameter, a; (d-f) lattice parameter, c; (c) ratio of lattice parameter, c/a, of ZnO film according to weight ratio of $H_2O$/solvent in the solution deposited by ESD at 300°C, 400°C and 500°C respectively.

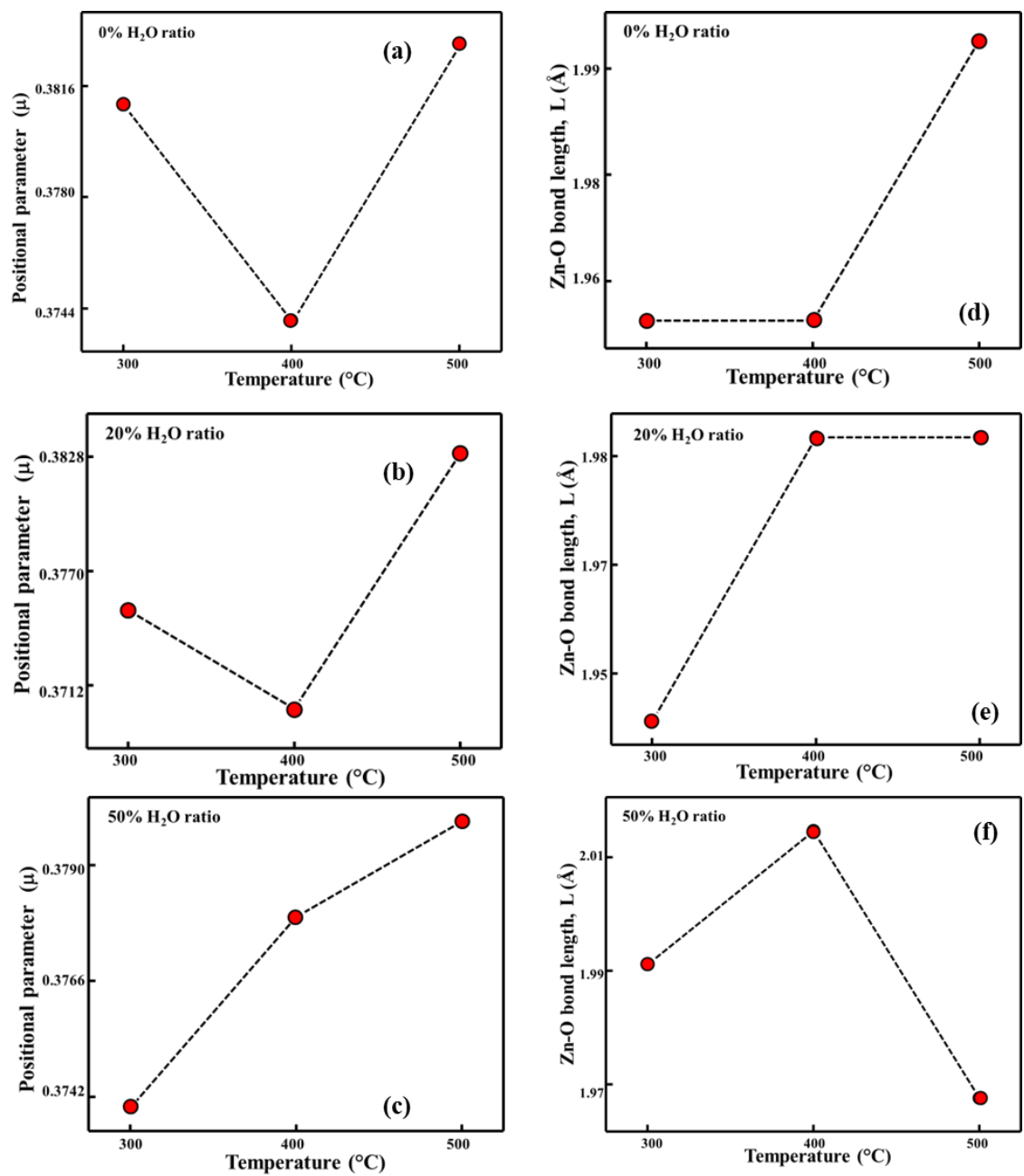

Fig. 4. The trend of (a-c) positional parameter, μ, and (d-f) Zn-O bond length of ZnO film according to weight ratio of H$_2$O/solvent in the solution deposited by ESD at 300°C, 400°C and 500°C respectively.

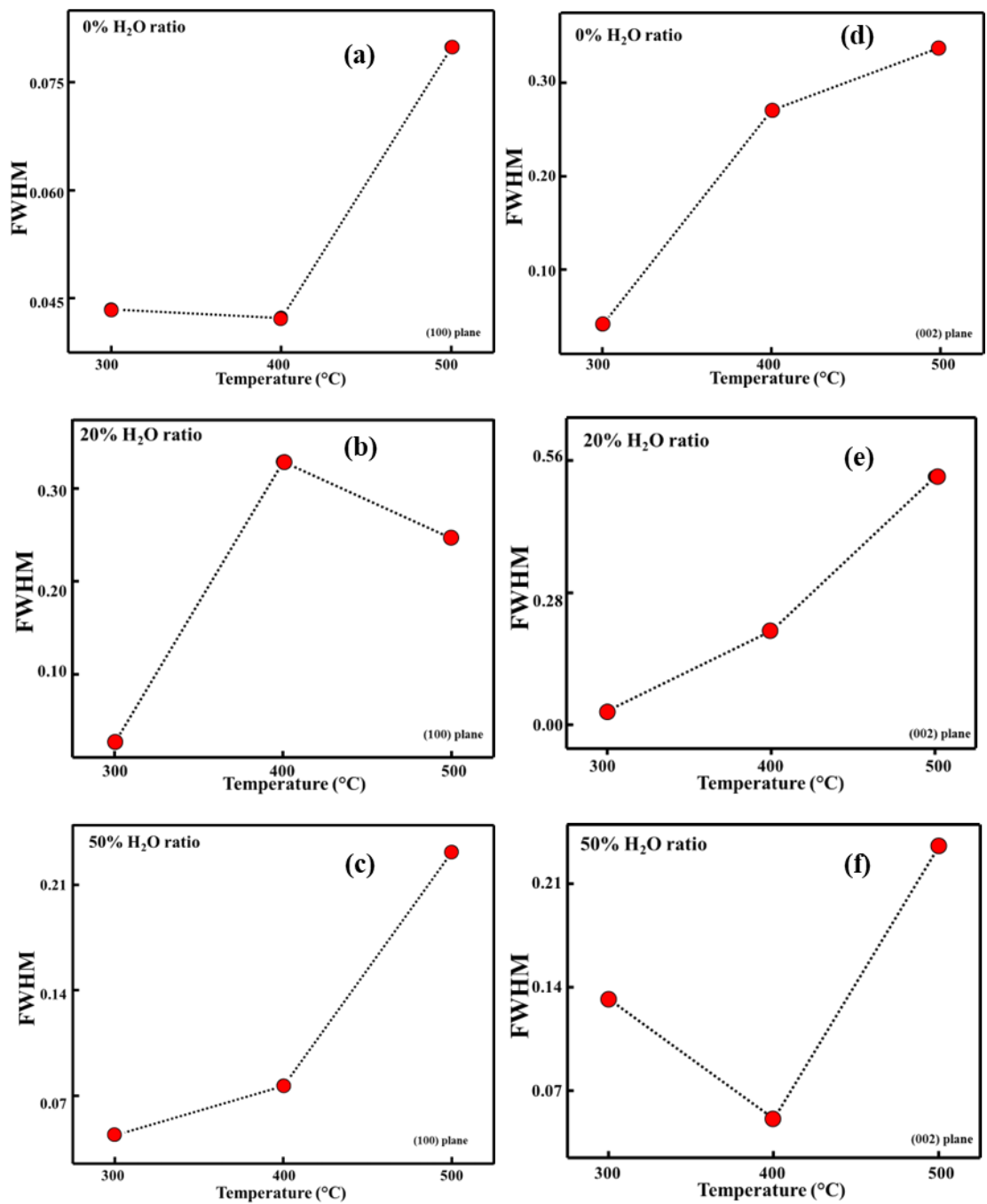

Fig. 5. FWHM for the (a-c) (100) plane, and (d-f) (002) plane of the XRD peak according to weight ratio of $H_2O$/solvent in the solution deposited by ESD.

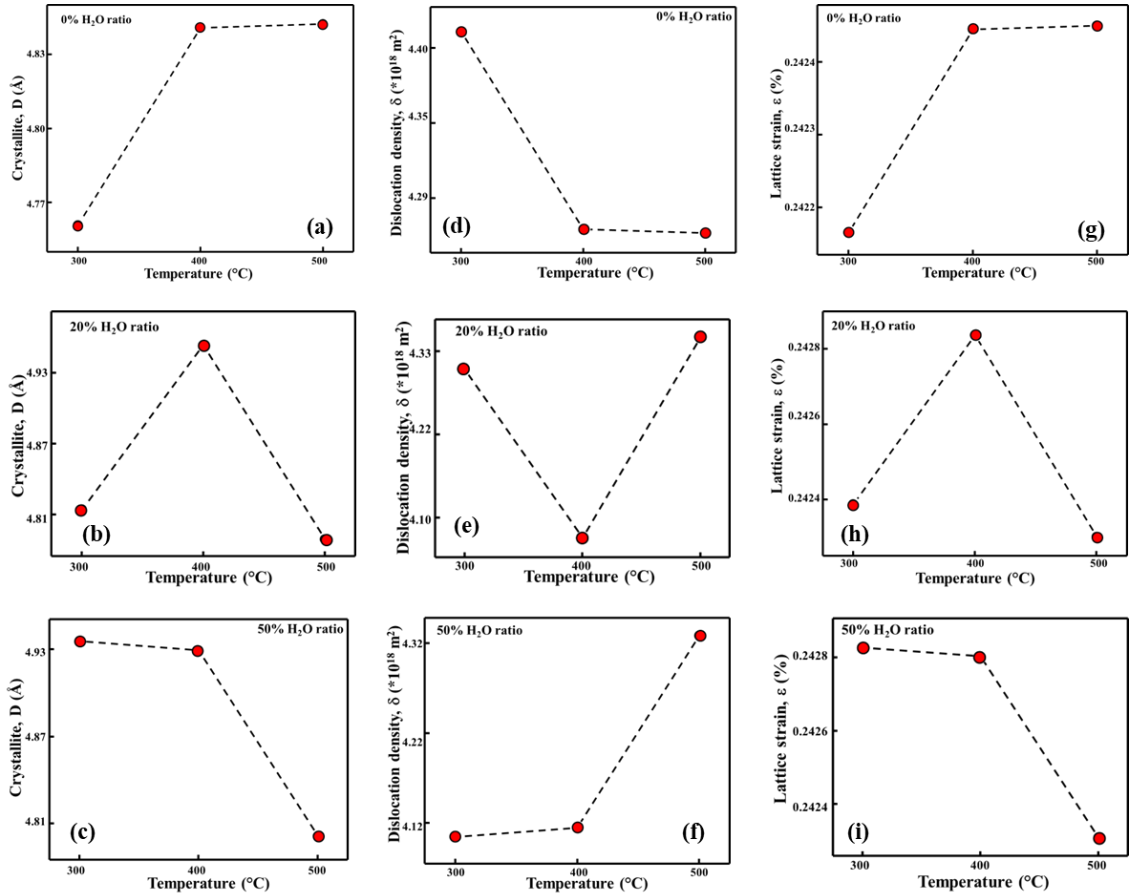

Fig. 6. The trend of (a-c) lattice crystallite, D; (d-f) Lattice dislocation density δ, (g-i) Lattice strain, ε of the XRD peak of ZnO film according to weight ratio of $H_2O$/solvent in the solution deposited by ESD at 300°C, 400°C and 500°C respectively.

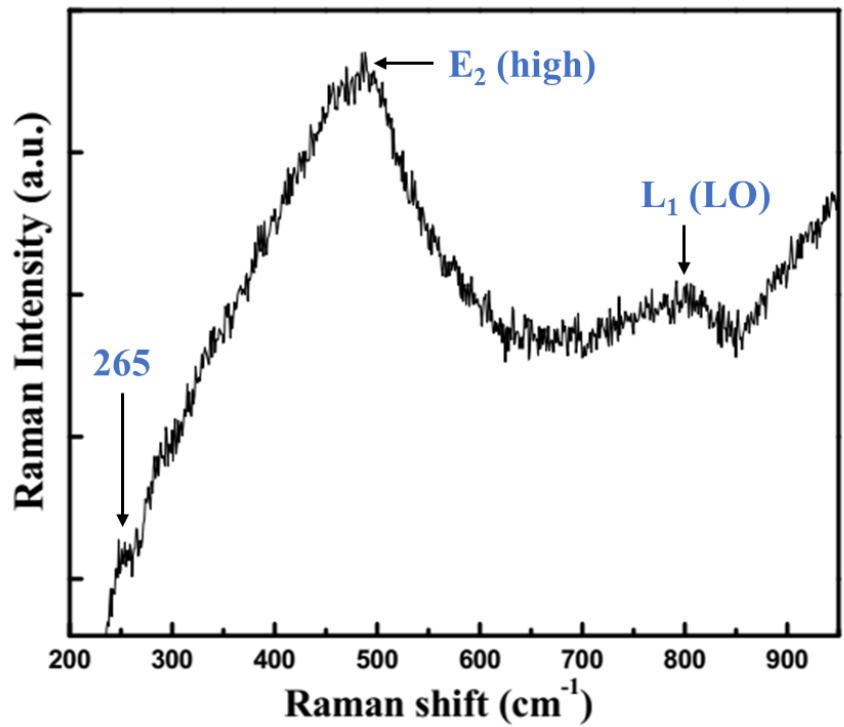

Fig. 7. Raman spectra of the ZnO thin films of $ZnCl_2$ and $CH_3CH_2OH$ precursor solution by ESD at 500°C temperature ESD sample with micro-ring structures.